\documentclass[]{spie}  

\usepackage[final]{graphicx}
\usepackage{subcaption}
\usepackage{siunitx}
\usepackage{amsmath,amsfonts,amssymb}
\usepackage{graphicx}
\usepackage[colorlinks=true, allcolors=blue]{hyperref}
\usepackage{cleveref}
\graphicspath{ {images/} }
\title{Realizing quantum nodes in space for cost-effective, global quantum communication: in-orbit results and next steps}
\author[a]{Chithrabhanu Perumangatt}%
\author[a,c]{Tom Vergoossen}%
\author[a,c]{Alexander Lohrmann}%
\author[a]{Srihari Sivasankaran}%
\author[a]{Ayesha Reezwana}%
\author[a]{Ali Anwar}
\author[a]{Subash Sachidananda}
\author[a]{Tanvirul Islam}%
\author[a,b]{Alexander Ling}
\affil[a]{
 Centre for Quantum Technologies, National University of Singapore, 3 Science Drive 2, S117543, Singapore
}
\affil[b]{Physics Department, National University of Singapore, 2 Science Drive 3, S117542, Singapore}
\affil[c]{Speqtral Pte. Ltd, 03-06 73 Science Park Drive Science Park 1, 118254,  Singapore   }

\authorinfo{Further author information:  Alexander Ling\\E-mail: alexander.ling@nus.edu.sg}

\begin{document} 
\maketitle

\begin{abstract}
Quantum sources and receivers operating on-board satellites are an essential building block for global quantum networks. SpooQy-1 is a satellite developed at the Centre for Quantum Technologies, which has successfully demonstrated the operation of an entangled photon pair source on a resource-constrained CubeSat platform. This miniaturized and ruggedized photon pair source is being upgraded to be capable of space-to-ground quantum key distribution and long-range entanglement distribution. In this paper, we share results from SpooQy-1, discuss their relevance for the engineering challenges of a small satellite quantum node, and report on the development of the new light source.
\end{abstract}


\section{INTRODUCTION}
\label{sec:intro}  
Deployment of quantum technology in spacecraft has been receiving greater attention over the past decade. Satellites that employ quantum systems in space have already enabled novel experiments to test fundamental laws of quantum physics\cite{xu2019satellite} \cite{BEC_ISS}, and demonstrated fundamental quantum communication protocols \cite{ren2017ground, dai2020towards}. These fundamental scientific experiments aside, the immediate application for space based quantum technologies is the global exchange of encryption keys via quantum key distribution (QKD). Pioneering experiments in the field have demonstrated the feasibility of QKD using satellites\cite{vallone15,tang2016generation,gunthner17,yin17,liao2017satellite,takenaka17,liao18, yin2020entanglement}. By integrating a satellite quantum node with large ground-based quantum networks, the first long distance integrated space and ground quantum network was recently presented\cite{chen2021integrated}. 

Despite these encouraging developments towards global quantum networks, a challenge that remains is the prohibitive cost of the technology. While high costs are already a challenge for commercial ground-based fibre quantum networks, satellite quantum nodes based on large satellites are orders of magnitude more expensive. For the space segment, a pathway to lower the costs is the use of smaller satellites. In particular standardized spacecraft such as CubeSats may play an important role in future quantum networks. Utilisation of small spacecraft as quantum nodes has progressed from conceptual design studies\cite{oi2017cubesat} to demonstrations of quantum technology \cite{tang2016generation}. 
An important requirement in both satellite and ground quantum networks is photonic entanglement. Finding cost-effective ways to generate, control and detect entanglement is imperative to enable practical quantum networks. The immediate application for generating and manipulating photonic entanglement on a satellite is Space-to-ground QKD in which the quantum correlations of entangled photon pairs are used to generate secure encryption keys, for example via the well-known BBM92 protocol\cite{bennett1992quantum}. Many different implementations of this QKD protocol have been published, some specifically targeted at space-to-ground QKD \cite{chapman2019hyperentangled, yin2020entanglement,ecker2021strategies}. The implementation of this protocol on a small spacecraft could significantly lower the costs for future satellite quantum nodes.

Very recently, we reported the successful operation of an entangled photon source on board the SPOOQY-1 CubeSat\cite{Villar:20}. In this paper, we share the latest experimental performance of SPOOQY-1 and discuss the next iteration of the quantum system.

\section{SPOOQY-1 Mission results} 

\begin{figure}[h]
  \centering
    \includegraphics[width=\textwidth]{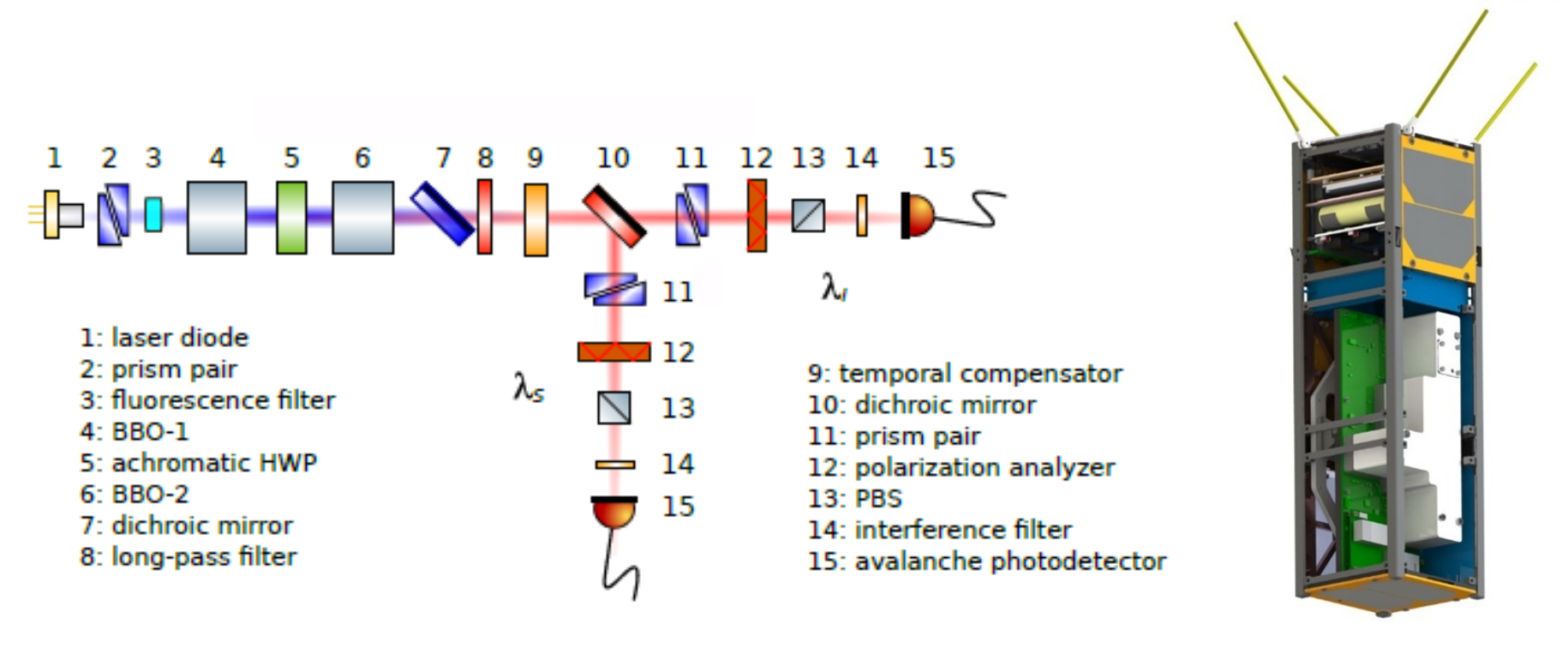}
    \caption{(Left) Optical layout of SPEQS-2. (Right) SpooQy-1 spacecraft with payload subsystems. The bottom two thirds contain the entangled photon source and the top third of the spacecraft contains the spacecraft avionics.}
    \label{fig:SPEQS2}
\end{figure}

The SpooQy-1 satellite demonstrated quantum entanglement correlations through a violation of the Bell-CHSH (Clauser-Horne-Shimony-Holt) inequality shortly after its deployment from the International Space Station on the 19th of June 2019\cite{Villar:20}. The satellite featured the second iteration of the Small Photon Entangling Quantum System (SPEQS-2), which improved upon the correlated source launched on Galassia \cite{tang2016generation} by the use of a different nonlinear crystal arrangement, and added optical components. The SPEQS-2 instrument generates polarization entangled photon pairs using an optical design in which the axes of the non-linear crystals (Beta Barium Borate : BBO) are aligned in a parallel configuration \cite{villar2018experimental}. This gives greater overlap for the orthogonally polarized photon pairs and offers high quality entanglement without the need for strong spatial filtering \cite{lohrmann2018high}. After splitting the entangled photon pairs into the signal and idler components, the two photon polarization correlation is characterized using Liquid Crystal Polarization Rotators (LCPRs), polarizing beam splitters and a pair of Geiger-mode avalanche photon detectors. \Cref{fig:SPEQS2} shows the optical layout within the SPEQS-2 instrument and a cut-out model of the satellite. Experiments on board SpooQy-1 continue as of February 2021 and the satellite is expected to de-orbit within the next 12 months.

\begin{figure}
    \centering
    \subfloat{\includegraphics[scale = 0.8]{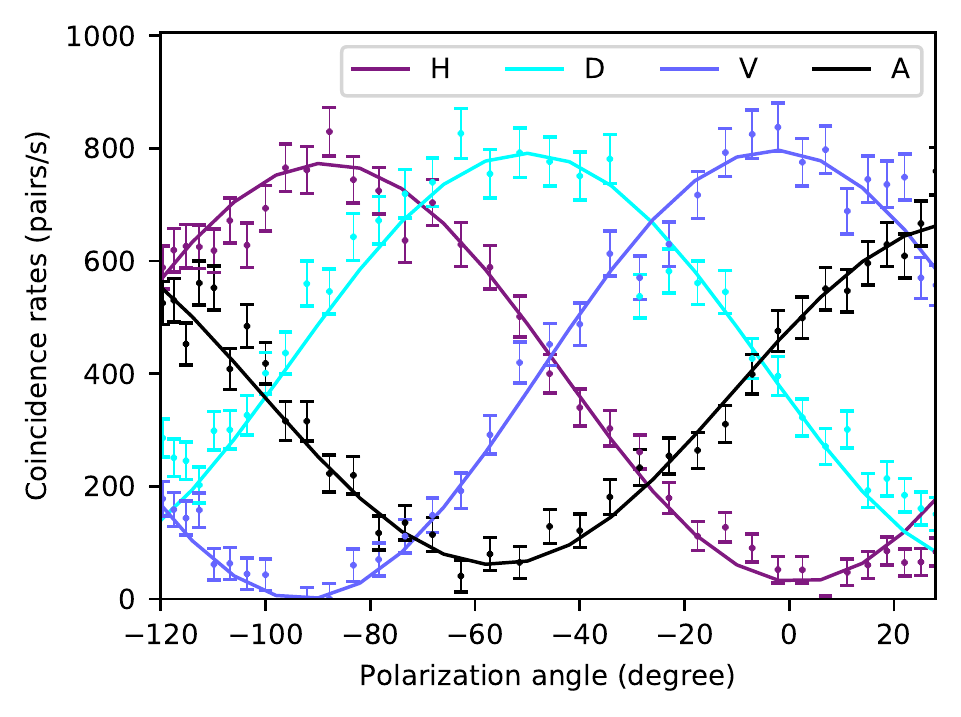}}
    \hskip -0.1ex
    \subfloat{    \includegraphics[scale = 0.9]{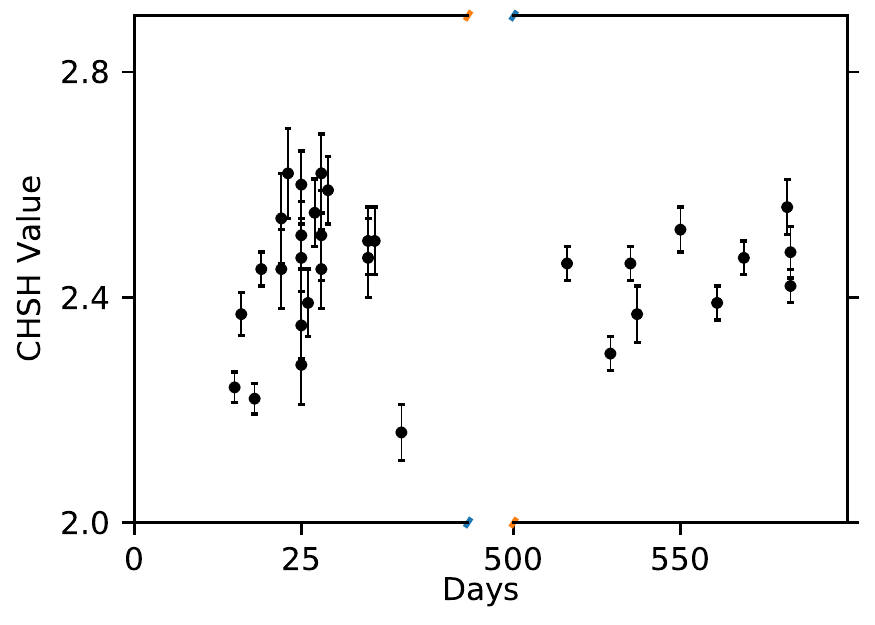}}

 \caption{(Left) Polarization measurement of the polarization entangled photon pairs produced by the payload. Plot reproduced from \citeonline{Villar:20}. (Right) CHSH value recorded within the first 50 days and 500-600 days after the deployment into orbit.}
    \label{fig:CHSH}
\end{figure}

The output of a typical entanglement test in orbit is shown in Fig.~\ref{fig:CHSH}. For this experiment, the LCPR in the idler path is set to fixed polarization measurement settings - horizontal, vertical, diagonal and antidiagonal (H/V/D/A). The LCPR in the signal path is scanned and the number of coincident detection events is recorded. The number of accidental correlations is estimated and removed. The expected sinusoidal behaviour ($C\propto cos^2(\theta)$ with $C$ denoting the number of coincidences and $\theta$ the measurement angle) is observed and fitted to the data. The $S$ parameter of the CHSH inequality is then estimated from the fit. For the measurement set in Fig.~\ref{fig:CHSH}(a), we estimate $S=2.53\pm0.06$ which is in good agreement with the pre-launch value recorded after payload assembly\cite{Villar:20}. The quality of entanglement as well as the reliability of the measurement setting are affected by the environment temperature and it is often necessary to heat the payload to a temperature range between \SI{15}{\celsius} and \SI{23}{\celsius} to achieve satisfactory entanglement quality. The quality of entanglement was monitored over the lifetime of the satellite and has not been observed to degrade, as shown in Fig.~\ref{fig:CHSH}(b).

In addition to the CHSH value, data collected over the past 595 days include temperature measurements within the payload and the spacecraft sub-systems and the detector dark counts for the two single photon detectors (SAP500). The dark counts are of particular interest, as displacement damage within the active area of the single photon APDs increases the dark count rate steadily as a function of total accumulated dose.

\begin{figure}
    \centering
    \includegraphics[scale = 1]{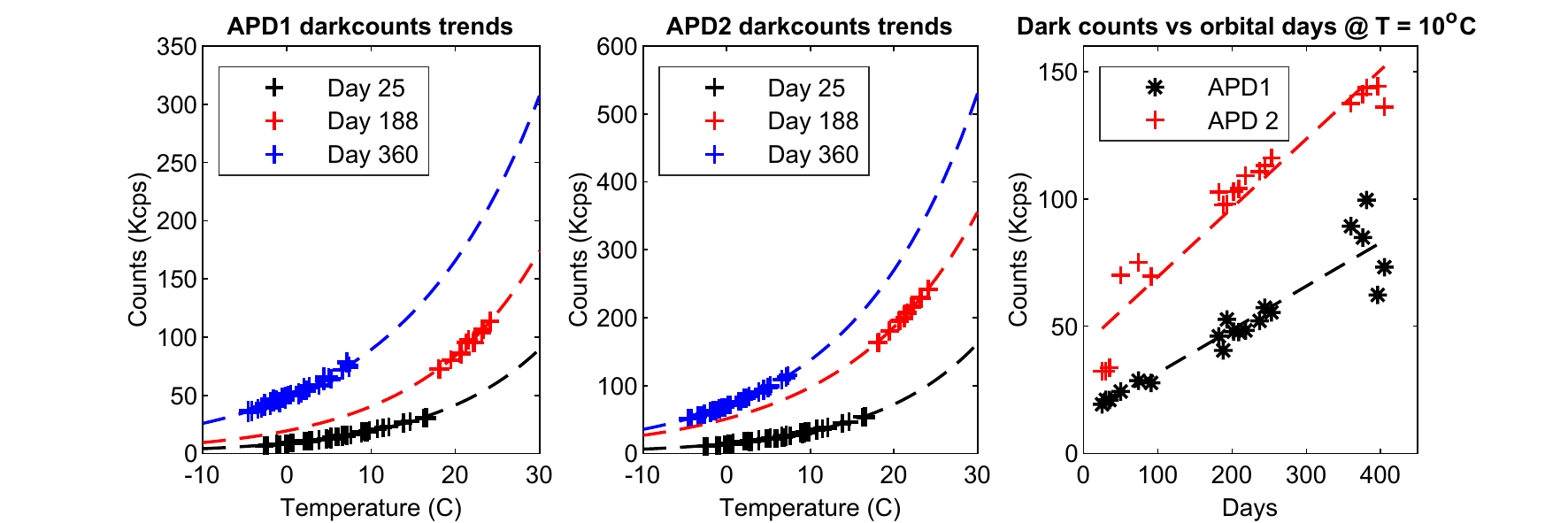}

 \caption{(a,b) Dark counts recorded over one orbit plotted as a function of temperature for 25, 188 and 360 days in orbit. The exponential fits are used to extrapolate a dark count value at a temperature of \SI{10}{\celsius}. (c) Dark counts for both detectors normalized to \SI{10}{\celsius} as a function of time spent in orbit. However, it is to be noted that the satellite was deployed 30 days after it was launched into a space environment (International Space Station).}
    \label{fig:darkcounts}
\end{figure}

\begin{figure}
    \centering
    \includegraphics[scale = 0.8]{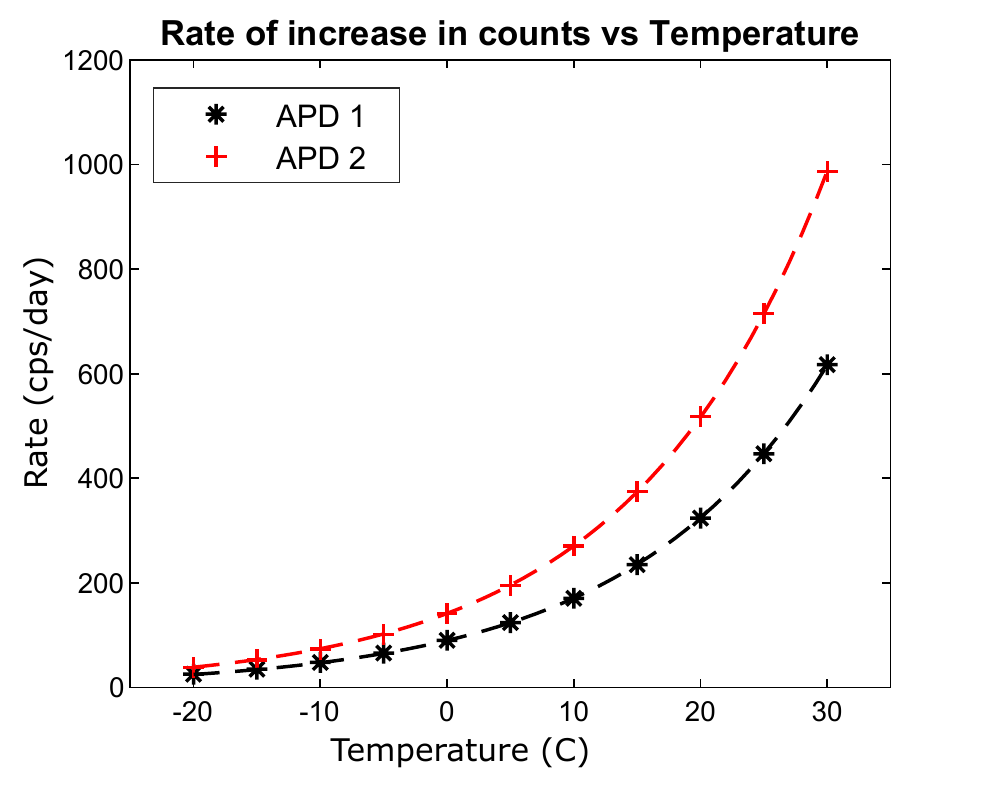}

 \caption{Change in the dark counts per day as a function of APD temperature. Each value for the rate are obtained from the fit given in \ref{fig:darkcounts}(c).}
    \label{fig:rate}
\end{figure}

Since the payload temperature and detector temperature is affected by the illumination time that changes throughout the year, it was not possible to record detector dark counts at a specific temperature consistently. Instead, the dark counts were logged throughout an orbit and plotted as a function of temperature. An exponential function that describes underlying thermal excitation of deep-level defects was fit to the dark counts. From the fit the dark count values at different temperatures were extrapolated. The dark count values at the beginning of the mission, after 188 days in orbit and after 360 days in orbit are shown in Fig.~\ref{fig:darkcounts}(a) and (b) for the two single photon detectors in the payload. The extrapolated dark count value for \SI{10}{\celsius} is shown in Fig.~\ref{fig:darkcounts}(c) as a function of days since deployment. The dark counts at this temperature increase by approximately 170~cps per day for the signal detector and  270~cps per day for the idler detector. Furthermore, the rate of change of dark counts per day is plotted against the temperature in Figure \ref{fig:rate}. Future missions will utilise cooled detectors to reduce the effects of radiation damage.

\section{Next generation quantum payload}
Building on the success of the SpooQy-1 mission, a next generation quantum payload was designed that is suitable for space-to-ground entanglement distribution. This section discusses the detailed design of the payload. In the first section, a QKD model is presented that was used to evaluate key performance parameters and to inform the design of the system. Next, the system components are introduced with detailed description of the individual subsystems. An experimental validation of the key components of the system is presented to confirm the model while validating the subsystem design and component choices. Finally, the design of a compact payload prototype is presented. 

\subsection{Performance modelling and design targets}
We begin by introducing the QKD performance model, discuss the validation steps and the most important findings.

Polarization entangled photon pairs are selected to implement the well-known BBM92 QKD scheme \cite{BBM92}. The polarization degree of freedom is chosen since photon polarization is readily preserved in free-space channels and it is simple and efficient to control and detect. In a first step, entangled photon pairs, said to consist of an idler and a signal photon, are created on the satellite. One photon is detected locally on-board the satellite while the other is transmitted to a ground receiver. The polarization of the signal and idler photons is measured in two orthogonal bases (e.g. horizontal-vertical: H-V and diagonal-antidiagonal: D-A). The measurement results of coincident detection events, i.e. detector \lq clicks\rq that occur within the same time window, are correlated if they corresponded to signal and idler photons from the same pair. Because of these non-classical correlations, once the spacecraft and the ground station have publicly compared which detector basis responded at a given time, they share common knowledge of the measurement results, from which a private encryption key can be extracted. Maximising the secret key means maximising the number of coincident detection events while minimizing the number of erroneous coincident detections. These erroneous event occur due to inherent imperfect entanglement quality, multiple photon pair emissions within the same coincidence time window, dark counts and background noise.   

We developed a detailed model for satellite-based QKD  \cite{VERGOOSSEN2020164} to estimate instantaneous/averaged asymptotic/finite key rates and Quantum Bit Error Rates for different scenarios. The inputs for the model as well as typical values for our design are given in \Cref{tab:Inputs}. Using the model a number of system trade off studies were performed.

\begin{figure}[]
  \centering
  \begin{minipage}[b]{0.49\textwidth}
    \includegraphics[width=\textwidth]{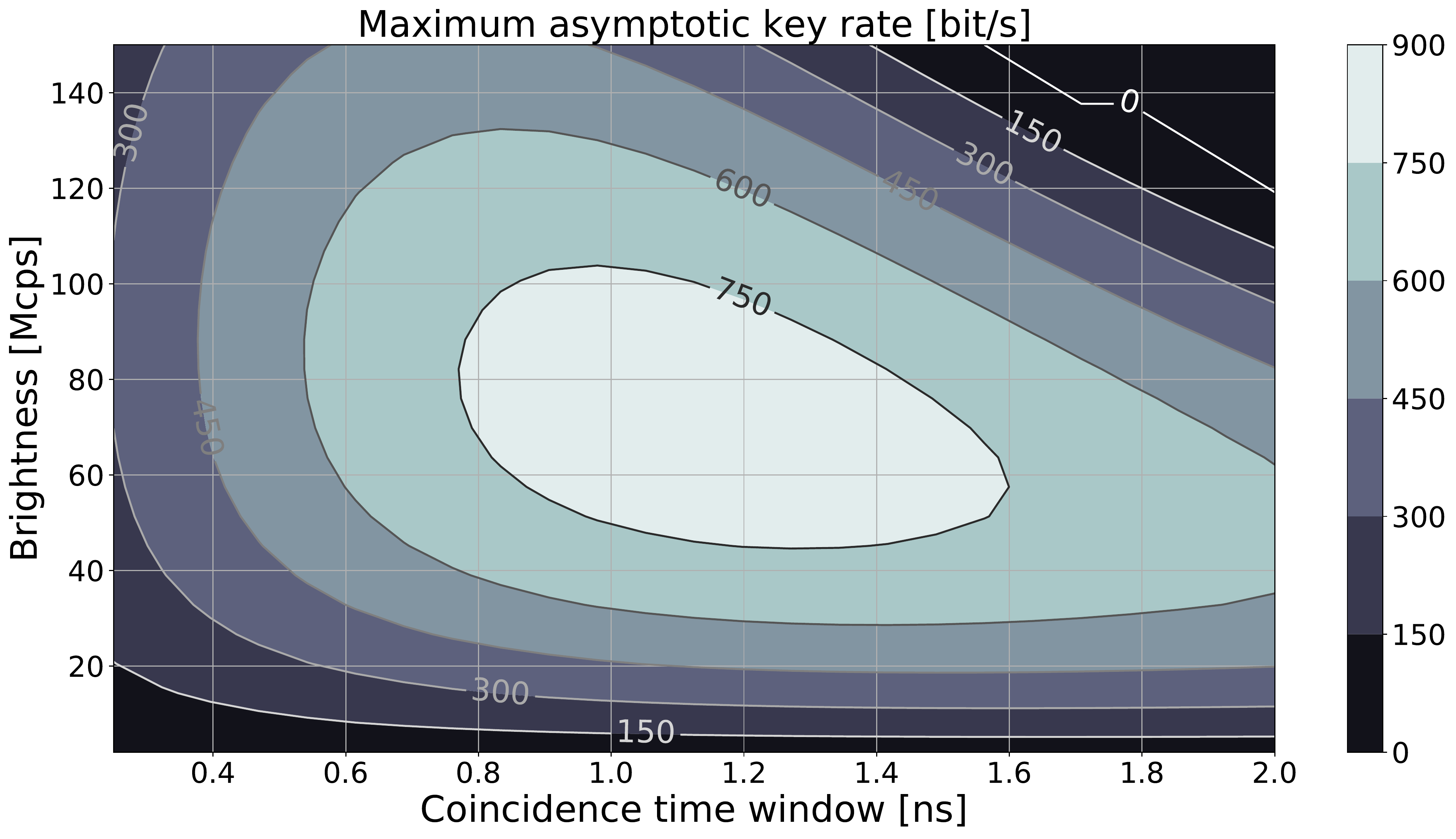}
  \end{minipage}
  \hskip -0.01ex
  \begin{minipage}[b]{0.49\textwidth}
    \includegraphics[width=\textwidth]{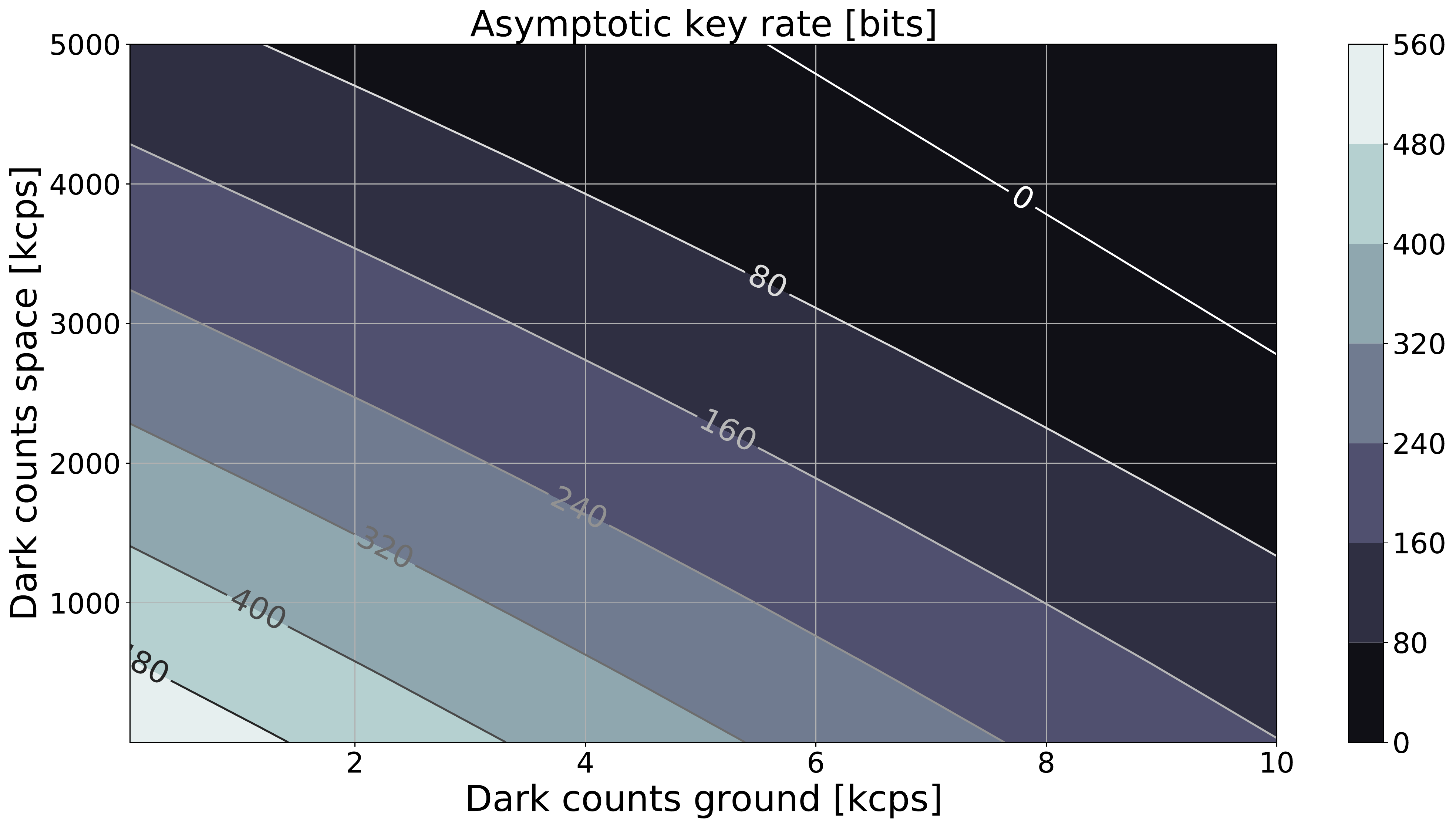}
  \end{minipage}
    \caption{(Left): Maximum asymptotic secret key rate for brightness and coincidence time window during a satellite pass straight over an optical ground station in good weather conditions at nighttime. (Right): Maximum asymptotic secret key rate for detector dark counts in space and on the ground during a satellite pass straight over an optical ground station in good weather conditions at nighttime. The brightness was set at 25MCps. Because of the high singles rates on the spacecraft the system is not sensitive to dark counts in space. On the ground, however, detector dark counts need to be limited. Input parameters as in \Cref{tab:Inputs}.}
    \label{fig:Rtau}
\end{figure}

First, a trade off of brightness (rate of photon pairs generated per second) and coincidence time window (see \Cref{fig:Rtau} (left)) shows that the optimum achievable asymptotic key rate (for the given input parameters) lies around 70M pairs per second at a coincidence time window of around 1-1.2ns. The product of the brightness and the coincidence window can be expressed as a mean photon pair number per coincidence window which in this case lies around 0.07. This is in good agreement with results obtained in long-distance free-space demonstrations for entanglement based QKD \cite{ecker2021strategies}.  

This optimum exists because accidental coincidences (detection of photons from two uncorrelated pairs within the same coincidence window) increase exponentially with the pair production rate rate, while coincident detection events only scale linearly, a fundamental limitation to the useful brightness of spontaneously generated entangled photon sources. At a 25\% pair to singles ratio, 70~M pairs per second would lead to 17.5~M photons detected per second in the space detection setup (4.375~M per detector). This exceeds the count rates of our time-stamping electronics currently qualified for space, which is around 8~M/s. However, a 1~ns coincidence time window is achievable given a detector jitter of around 300~ps for common single photon avalanche diodes. This implies that time transfer between the space and ground detection setups must be performed at better than 1~ns and the accuracy with which each detection event is measured should be at least 0.5~ns or lower. 

Second, a trade off of detector dark counts in \Cref{fig:Rtau} (right) illustrates a relative insensitivity of asymptotic key rate to detector dark counts in space and a strong sensitivity to dark counts on the ground, due to the low chance of correlating a noise event on the satellite with a detection event on the ground station, but the high chance of correlating a noise count on the ground with one on the spacecraft. The latter conclusion also applies to limiting background light, which can be done through spectral filtering (limited by SPDC bandwidth), temporal filtering (coincidence time window limited by detector jitter) and spatial filtering (reduction of field of view of receiving telescope is limited by pointing performance of the telescope and tracking mount). This becomes a key performance requirement for the optical ground station. On the spacecraft dark counts should be limited such that they do not impact the performance of the system, including time-stamping capacity.  Radiation damage will increase dark counts, probably more severely than on SPOOQY-1 due to the higher altitude, near-polar orbit of useful QKD satellites. Mitigation techniques include cooling of the detectors and thermal annealing of the active area of the detectors. Both options are planned for the next spacecraft.

\begin{figure}[]
  \centering
  \begin{minipage}[b]{0.49\textwidth}
    \includegraphics[width=\textwidth]{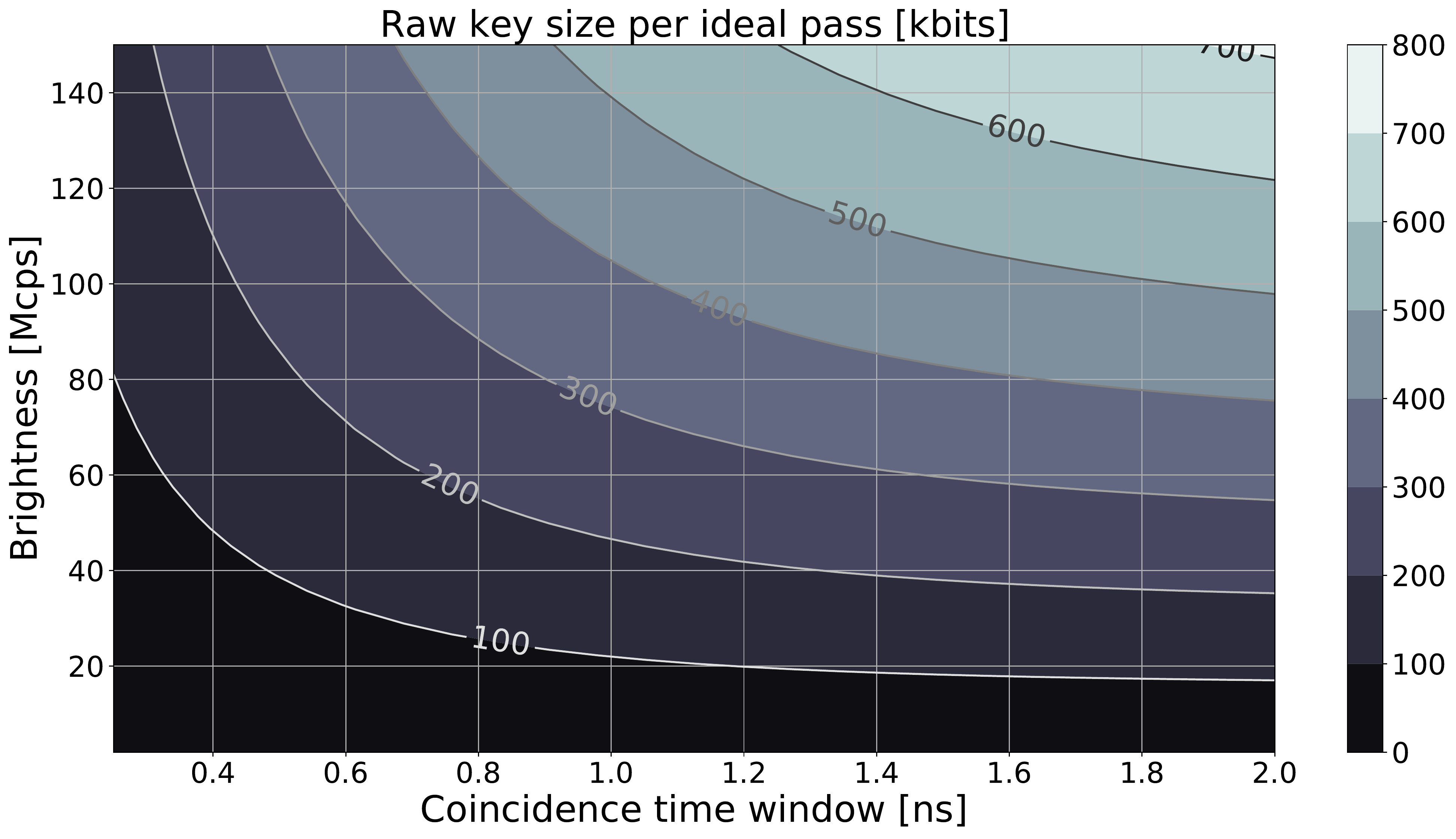}
  \end{minipage}
  \hskip -0.1ex
  \begin{minipage}[b]{0.49\textwidth}
    \includegraphics[width=\textwidth]{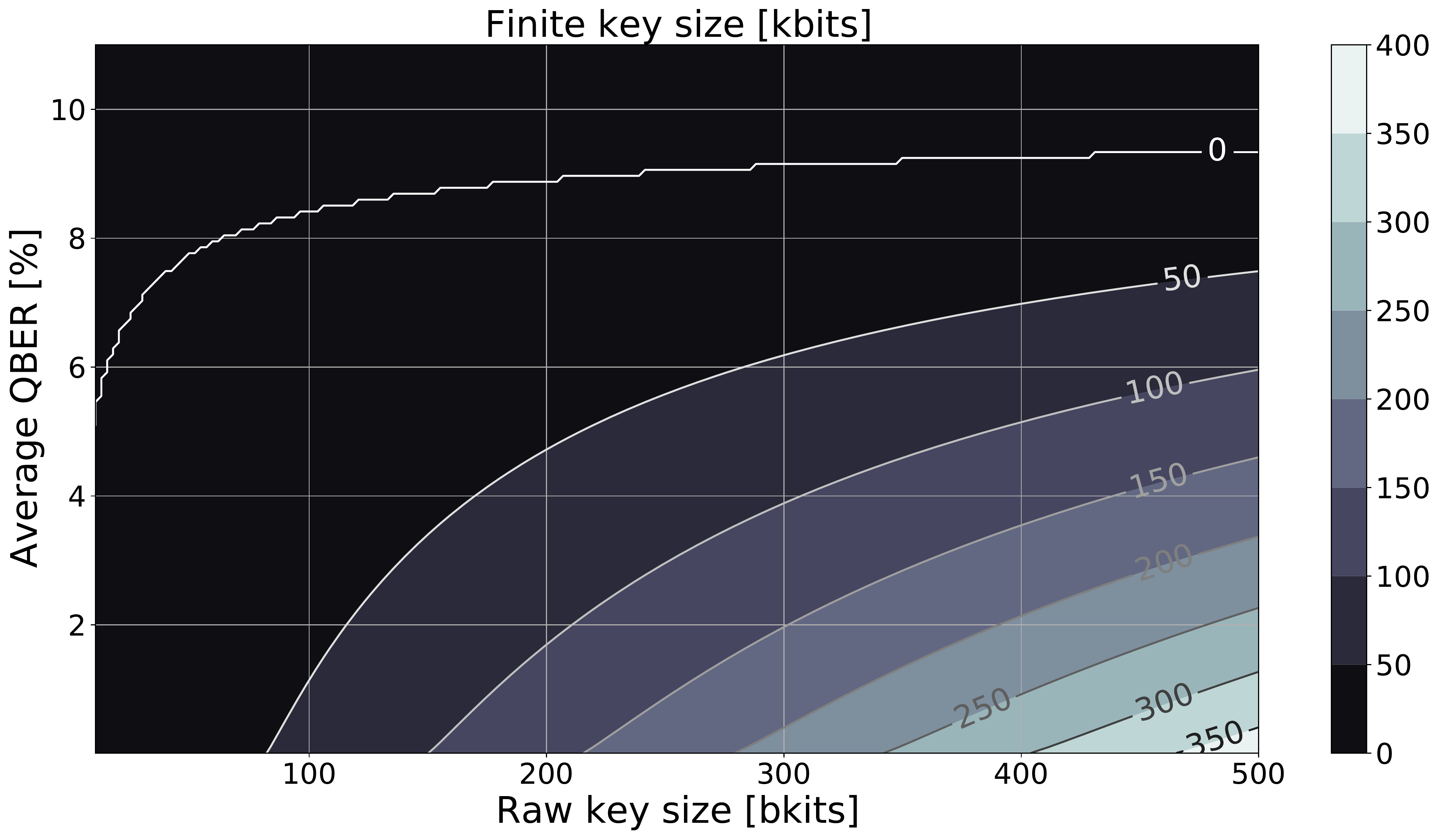}
  \end{minipage}
    \caption{(Left) Raw key created during one ideal pass. (Right): Finite key size for a given raw key size and average QBER as calculated from Appendix A.2 in \cite{chapman2019hyperentangled}. The security parameter $\epsilon_{cor}$ (the probability that Alice and Bob’s key are non-identical) was set as $10^{-10}$ and $\epsilon_{sec}$ (the probability that the adversary’s total quantum information is statistically distinguishable from the ideal output state) was also set as $10^{-10}$. Other parameters as in \Cref{tab:Inputs}.}
    \label{fig:finite}
\end{figure}

\begin{figure}[]
  \centering
  \begin{minipage}[b]{0.49\textwidth}
    \includegraphics[width=\textwidth]{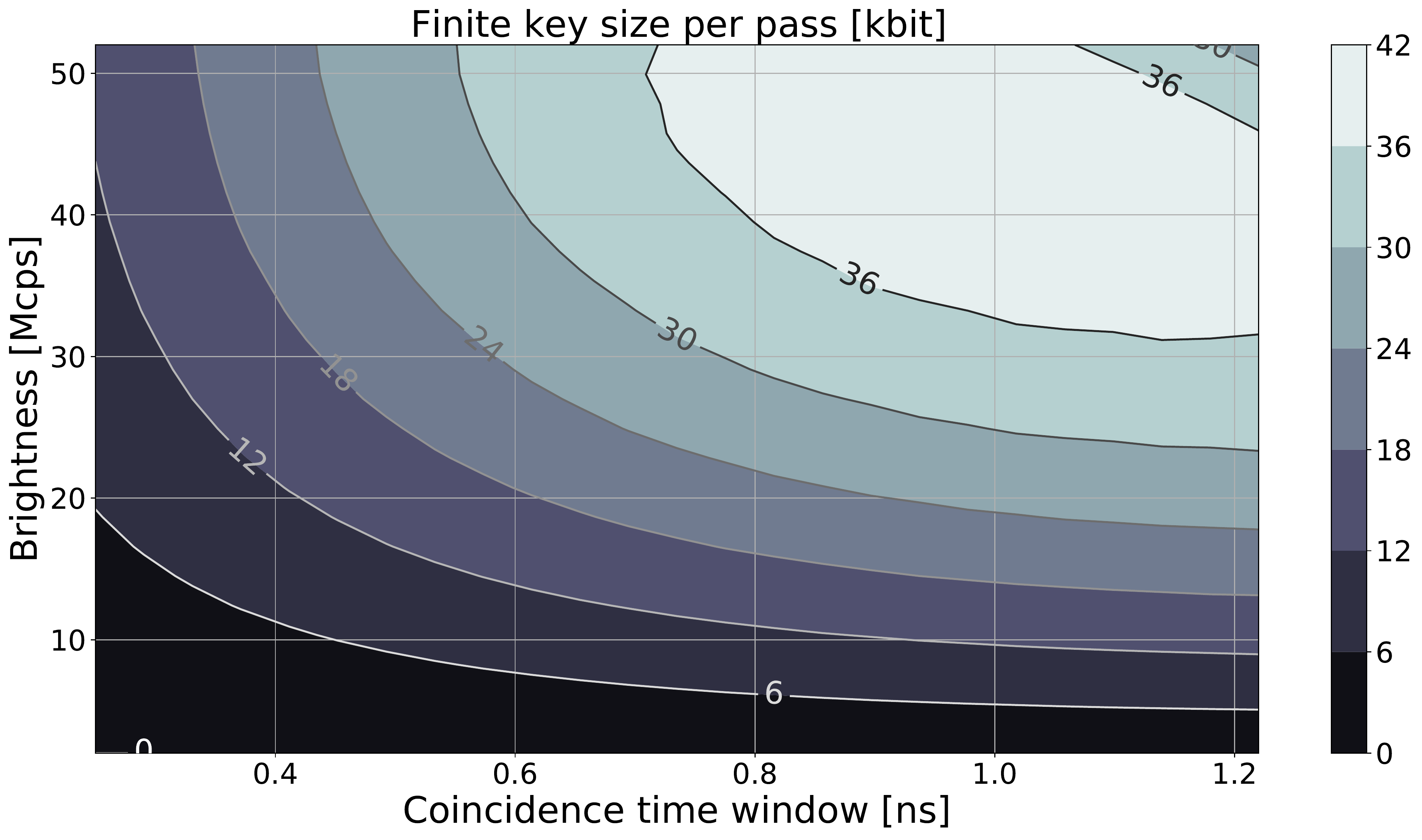}
  \end{minipage}
  \hskip -0.1ex
  \begin{minipage}[b]{0.49\textwidth}
    \includegraphics[width=\textwidth]{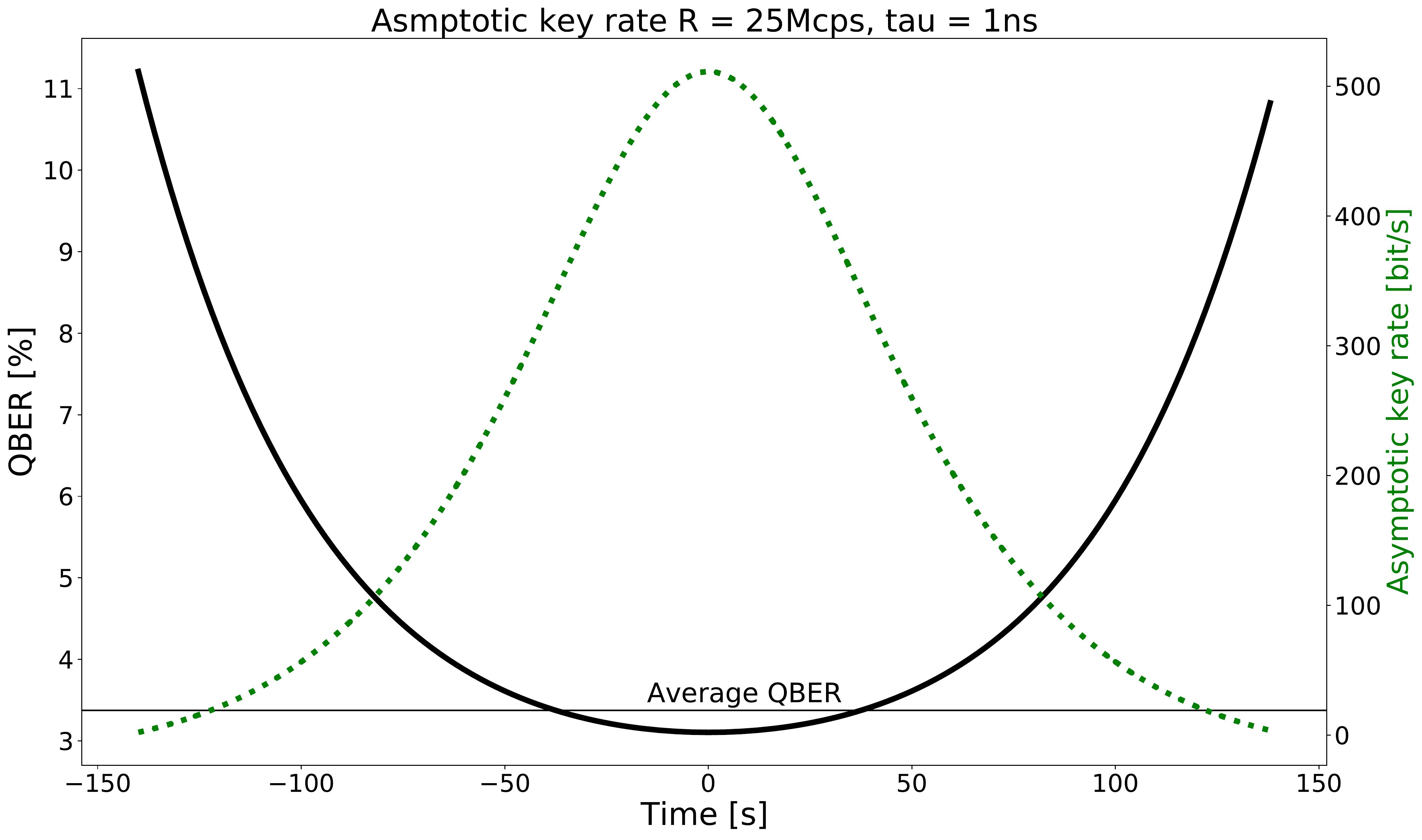}
  \end{minipage}
    \caption{(Left) Finite key size for one pass (same parameters as \Cref{fig:finite}), but the instantaneous QBER is limited to 5\% (so the average QBER is actually slightly below this). (Right) QBER and asymptotic key rate during one pass for a brightness of 25Mcps and a coincidence time window of 1ns. Other parameters as in \Cref{tab:Inputs}.}
    \label{fig:finite5}
\end{figure}

Another consideration is the system Quantum Bit Error Rate (QBER), which comprises contributions from the polarization visibility of the source, multi-photon pair events, dark counts, noise counts, polarization reference frame error and polarization state mixing introduced by elements in the optical link. From an engineering perspective, it is useful to manage this through a QBER budget, where individual contributions must be allocated. To set a QBER limit, instead of considering the asymptotic key rate, we can evaluate the effect of finite raw key sizes on the final secret key. The model calculates the finite key size for the total raw key accumulated during a pass following a recent security bound for the BBM92 protocol \cite{chapman2019hyperentangled}. First, the size of the raw key (i.e. sifted coincidences, before error correction and privacy amplification) that can be created during one ideal pass is shown in \Cref{fig:finite} (left). Achievable raw key sizes are between 100 and 300~k bits per pass for a brightness between 25 and 75~M pairs per second at a coincidence time window of 1~ns. \Cref{fig:finite} (right) shows that the total QBER should be limited to below about 8\% to get a positive finite key and to between 1-2\% to lose only 50\% during finite key extraction. From an operational perspective there is always a possibility of accumulating raw key over multiple passes, which can mitigate higher average QBERs. A tentative average QBER target can be set at 5\%, although during operations the average QBER can possibly be optimized by discarding sequences with high QBER. The finite key size, as well as instantaneous asymptotic key rate and QBER that are achievable within one pass are shown in \Cref{fig:finite5}. A QBER budget with targets for the individual contributions is shown in \Cref{tab:QBERbudget}.

\begin{table}[]
\caption{QBER budget for a space-to-ground QKD system.}
\label{tab:QBERbudget}
\begin{tabular}{|l|c|c|}
\hline
\textbf{QBER contribution} & \textbf{Effect on QBER} & \textbf{Target} \\ \hline
Source visibility: QBER=(1-VIS)/2 & Constant & 1\% \\ \hline
Detector dark counts (space) & Varies with coincidence rate & Limit dark counts to \textless{}100kcps \\ \hline
Detector dark counts (ground) & Varies with coincidence rate & Limit dark counts to \textless{}1kcps \\ \hline
Background light & Varies with coincidence rate & Limit background light to \textless{}1000cps \\ \hline
\begin{tabular}[c]{@{}l@{}}Total noise counts\\  (dark counts and background counts)\end{tabular} & Varies with coincidence rate & \textless{}1\% at closest approach \\ \hline
\begin{tabular}[c]{@{}l@{}}Multi-photon pairs in one \\ coincidence time window \\ leading to an accidental coincidence\end{tabular} & Constant & \begin{tabular}[c]{@{}c@{}}\textless{}1.5\%, but optimize for this\\  by choosing ideal brightness\end{tabular} \\ \hline
\begin{tabular}[c]{@{}l@{}}Polarization reference frame rotation \\ correction error\end{tabular} & Constant & 0.1\% \\ \hline
Polarization state mixing & Constant & 0.1\% \\ \hline
\textbf{Total QBER} & \multicolumn{2}{c|}{\textbf{$\sim$3.5\% at closest approach, $\sim$5\% on average}} \\ \hline
\end{tabular}
\end{table}

\subsection{Description of sub-systems}
\begin{figure}
    \centering
    \includegraphics[scale=0.35]{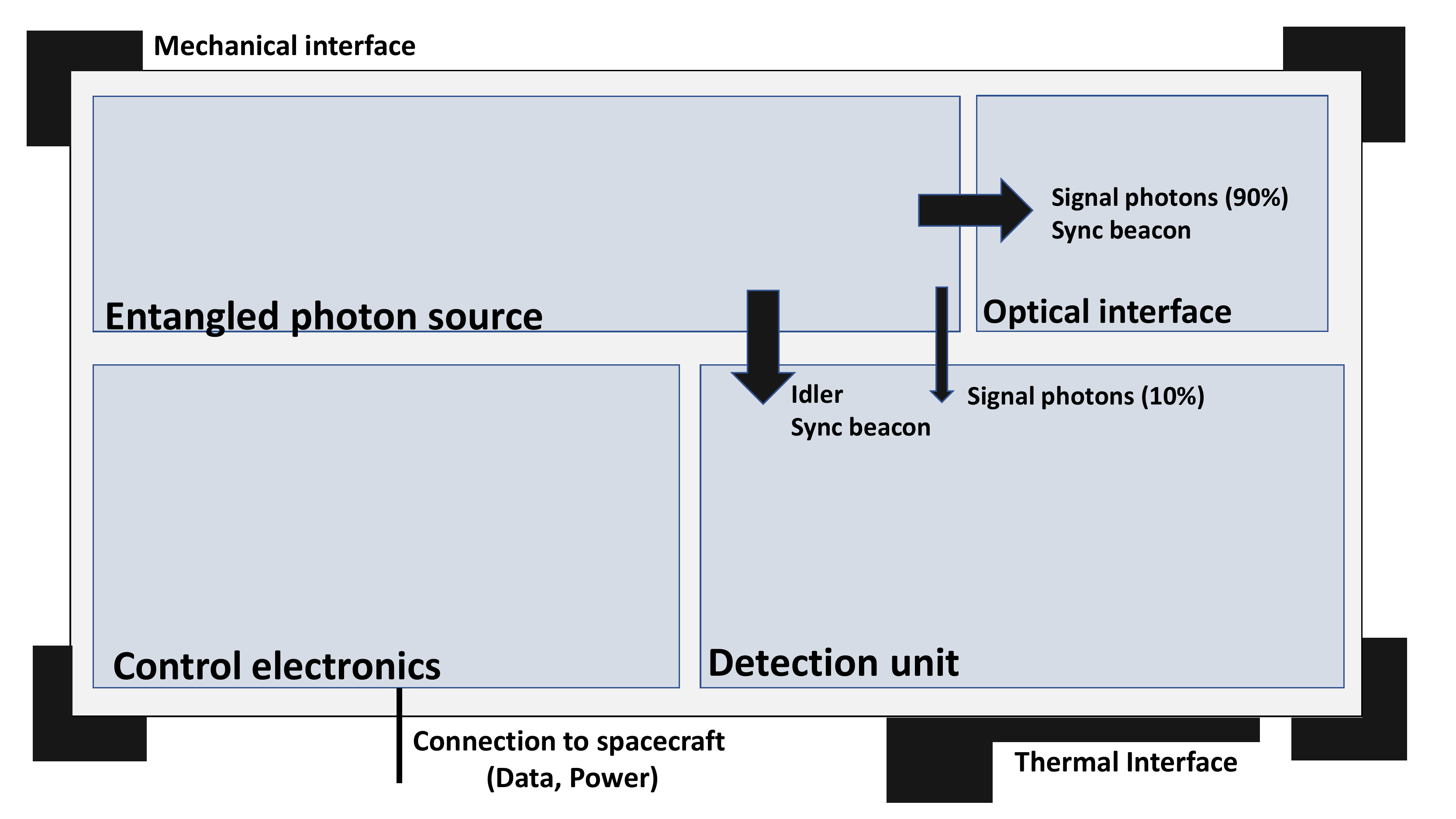}
    \caption{Physical architecture of the quantum payload.}
    \label{fig:PhysArch}
\end{figure}

\begin{figure}
    \centering
    \includegraphics[scale=0.7]{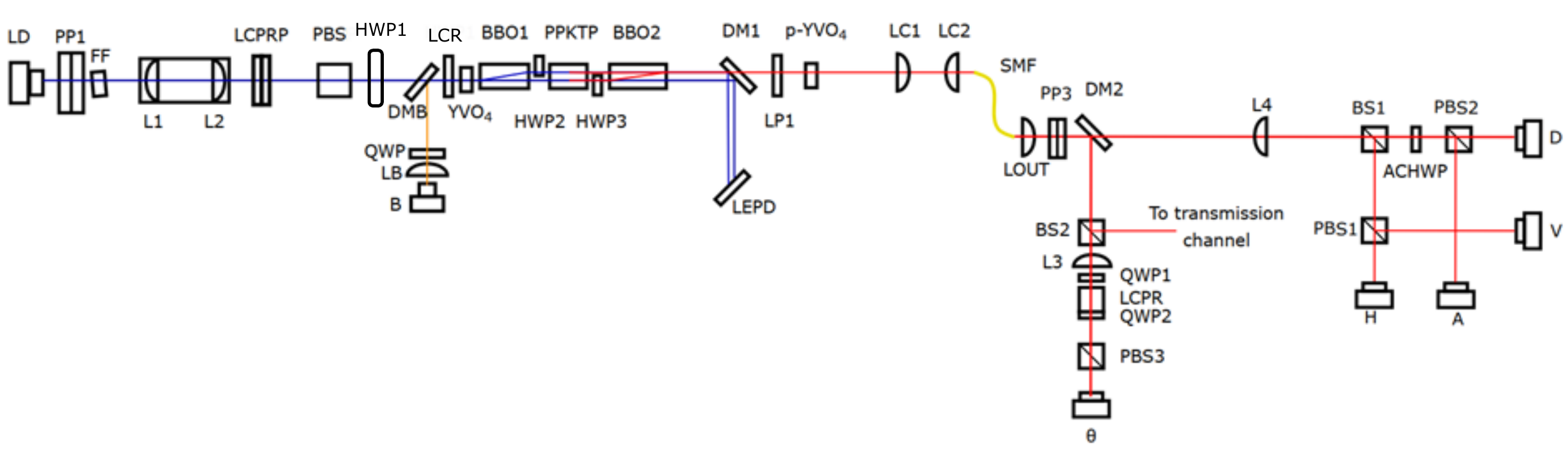}
    \caption{Schematic of the quantum payload. LD- laser diode, pp - prism pair, LCPR - liquid crysat polarization rotator, PBS - polarizating beam splitter, DM -B beacon dicroic mirror, L- lens, B- beacon laser diode, QWP- quarter wave plate, HWP- half wave plate, LB- beacon lens, $YVO_4$ - yttrium orthovanadate , BBO - beta barium borate,DM- dichroic mirror, LP- long pass filter, LEPD- lateral effect photodiode,   LC - coupling lens, SMF- single mode fibre,LOUT- output lens, BS- beam splitter, ACHWP- achromatic half wave plate and H,V,D,A, $\theta$ - APDs.  }
    \label{fig:full}
\end{figure}

The quantum payload consists out of an entangled photon source, a detection unit (Alice), control electronics and software, an optical interface and conventional spacecraft interfaces. This is illustrated in \Cref{fig:PhysArch}. The source is based on parametric down conversion producing non-degenerate pairs of photons, signal and idler, entangled in the polarization degree of freedom. Figure \ref{fig:full} describes the optical layout of the quantum payload. Within the quantum payload the signal (785~nm) and idler (837~nm) photons, are separated using a dichroic mirror. The signal is routed to the telescope payload using an interface mirror which transmits it to the ground station while the idler photons are measured using a 4-channel polarization detection unit that projects incoming photons at random into the H/V or A/D bases. To monitor the efficiency of the coupling of entangled photons, as well as the quality of the generated entangled state, we employ a ``health check detector" analysing the polarization state of a small portion of the signal photons. Along with the entangled photons we are transmitting a timing beacon laser that will be used for time synchronization.  

\subsubsection{Source}
Entanglement in polarization is achieved by superposing orthogonally polarized  SPDC photons generated using two pump decay paths \cite{anwar2020entangled}. Most designs for high performing entangled sources make use of a Sagnac interferometer. In this case, the phase stability of the entangled state generated is related to the alignment stability of the mirrors used in the interferometer with counter propagating pump beams. Moreover, Sagnac designs suffer from a relatively large footprint that is required for the interferometer. Therefore, instead of the common Sagnac source, we use a liner displacement interferometer based design in which two displaced pump beams co-propagate inside a second order non-linear crystal. Such linear designs can be more compact and are particularly easy to align \cite{lohrmann2020broadband}. 

The source concept is briefly reiterated here. The pump beam of diagonal polarization is split into a horizontal and vertical beam using spatial walk off from a birefringent crystal. The horizontally polarized pump beam is converted to vertical polarization using a half wave plate. A periodically poled potassium titanyl phosphate crystal kept at a constant temperature inside an oven is used as the second order non-linear crystal for SPDC. The crystal is oriented such that the vertically polarized pump photon with 405 nm wavelength is down converted to two vertically polarized daughter photons with wavelengths 785~nm (signal) and 837~nm (idler). The polarization of one of the pump decay path is converted to horizontal and the two decay paths are superposed using another birefringent crystal of appropriate length.

Propagation of SPDC photons through the birefringent crystal can a introduce wavelength dependent phase between the orthogonal states. This can reduce the entanglement visibility which in turn increases the quantum bit error rate. To compensate for the wavelength dependent phase a  Yittrium Orthovanadate (YVO4) crystal is introduced. We also introduce a pre-compensator crystal in the pump beam path, with which the phase introduced by different pump wavelength is always constant. This is helpful to ensure that the generated state is resilient to the changes in the pump laser wavelength that are generally associated with volume holographic grating laser diodes. A liquid crystal retarder (LCR) is used to control the relative phase between horizontal and vertical pairs.

The pump beam is removed using a dichroic mirror and a long pass filter. The SPDC photons are coupled into a single mode fiber. Coupling the entangled photons into a single mode fiber is the most challenging part of the payload development. To minimize the footprint of the instrument, as well as to avoid mirror mounts with spring mechanisms, we use lens translation to couple the beam into the fibre. L2 is translated along the beam path (z direction) while L1 and LC1 are translated perpendicular to the beam propagation direction (X and Y). The distance between L1 and L2 changes the pump focus at the crystal. The pump and collection focus are optimized in order to optimize the efficiency of collection as well as the brightness \cite{dixon}, with pump waist sizes around \SI{100}{\micro\metre} and a collection waist of \SI{55}{\micro \metre}.  We use a single mode fiber of short length (100 mm) with collimators attached to both ends (LC2 and LOUT). 

The source also includes the timing beacon that needs to be transmitted along with the quantum signal. This is a 785~nm pulsed laser with frequency  between 1~kHz to 50~kHz (tunable) and a pulse width of 5 ns. A portion of this laser is transmitted through the dichroic mirror (DM2) causing a 4-fold coincidence on the on-board detection system. Most of the beacon is transmitted along with the signal photons to the ground station. The 4-fold coincident clicks corresponding to the beacon can be extracted directly from the timestamps and are used to correct any time drifts (Doppler or clock drifts) between the satellite and the ground stations by cross correlating beacon events. This correction is fast and robust up to clock drifts of several \SI{100}{\micro\second}.

\subsubsection{ Detection unit}
One photon of the entangled pair is detected within the satellite. A Dichroic mirror (DM2) transmits the signal and reflects the idler photons. The detection setup is a standard 4-channel polarization measurement device.  A 50:50 beam splitter (BS) selects the measurement basis (H/V or D/A) at random while an achromatic wave plate (ACHWP) and polarizing beam splitter (PBS) project the photons into two orthogonal polarization states for each basis choice. A lens (L2) is used to focus the idler photons to the active area of the detectors. Free space detectors with an active area of 180 $\mu m$ with in-built thermo-electric cooler are used. The detectors are actively quenched using home-built fast quenching circuits with a dead time of less than 100~ns. The photon arrival time is recorded by a time stamp with resolution better than 0.5~ns. 

A part of the signal that is reflected from  DM2 is sent to the health check detector. A 90:10 beam splitter is used for transmitting a part of the signal to the health-check detector. The coincidence counts between this detectors and the other 4 detectors measure the pair to singles-ratio and source visibility (QBER). The error signal produced from the coincidences is used to maintain an appropriate phase of the quantum state by applying a corresponding voltage to the LCR. Additionally, the coincidence rates are used to control the laser power and the temperature set points of the thermal control loops of the laser and the nonlinear crystal. 

\subsubsection{Optical interface}
The quantum payload is designed such that the payload will be mounted back to back with the optical terminal payload. This free-space optical interface between the payloads will involve an interface mirror reflecting the signal and the beacon from the quantum payload to the optical terminal payload. The 1/$e^2$ diameter beam diameter of the quantum and synchronization beams is 3.5~mm. 

\subsection{Validation of key rate model and design}
\begin{figure}[]
    \centering
    \includegraphics[scale=0.8]{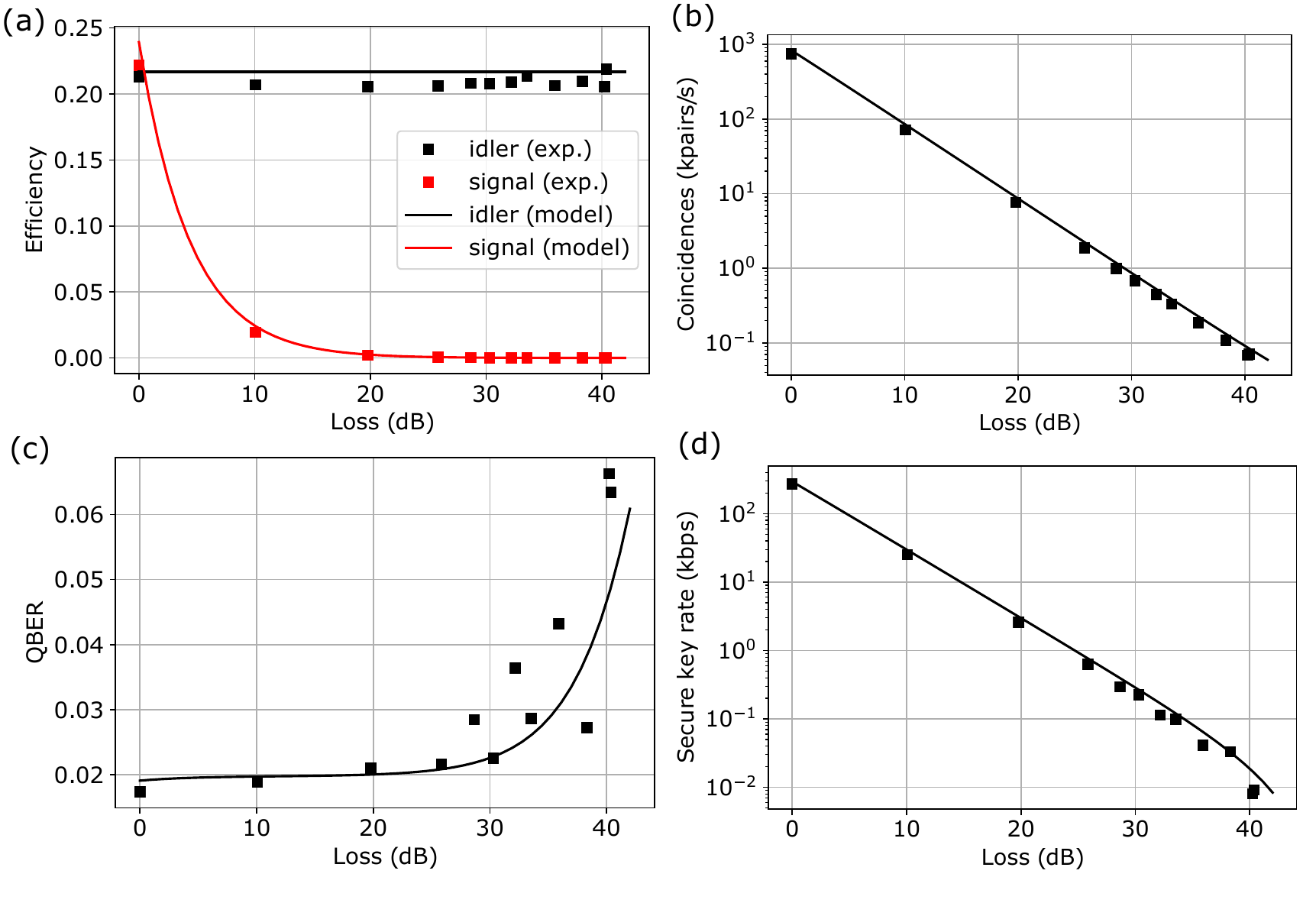}
    \caption{ Table top verification of the link model. The filled squares are the experimentally obtained data and the lines are from the model. Variation of (a)  heralding efficiency for signal and idler photons  (b) Detected coincidence  (c) variation of quantum bit error rate and  (d) estimated secret key rate are plotted against the introduced channel loss.  }
    \label{fig:TTD}
\end{figure}
\begin{figure}[]
    \centering
    \includegraphics[scale=1]{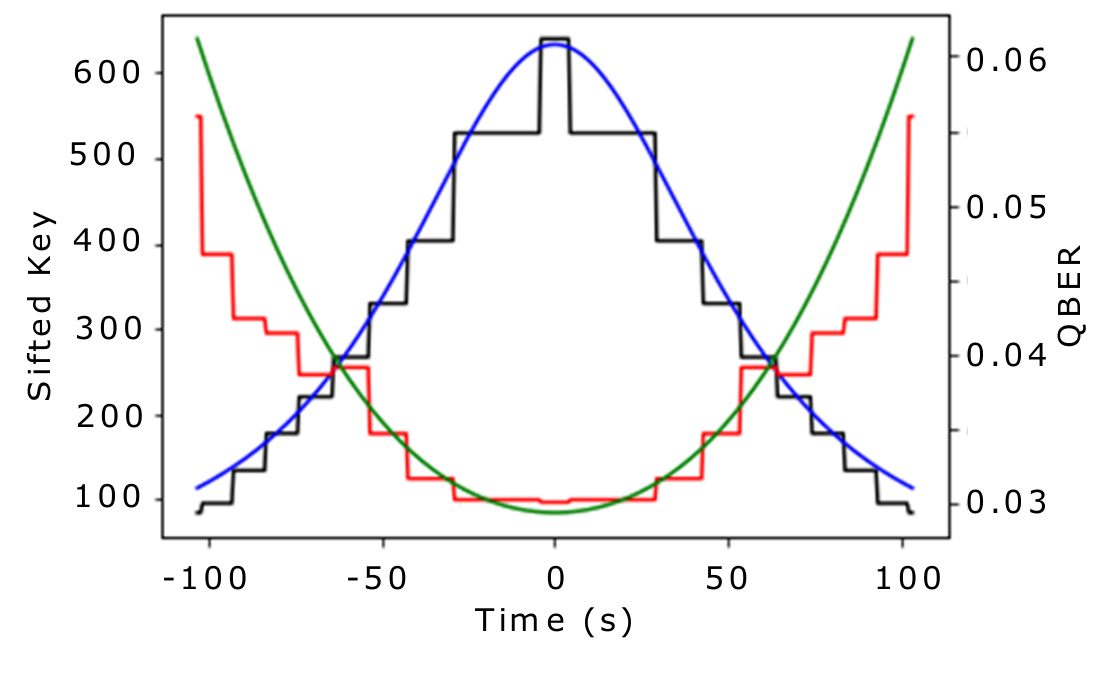}
    \caption{ A simulated satellite pass and the performance of the table top system. The black and blue lines are the measured and predicted sifted key. Similarly, red and green curves are the measured and predicted QBER. Initial parameters:- source QBER: 0.9\%, average efficiency : 22\%, generated pair rate: 34 Mcps, Coincidence window: 1.5 ns.    }
    \label{fig:pass}
\end{figure}

We implement the above mentioned experimental setup for  QKD on an optical table. The key rate calculation as well as the optical design were experimentally validated using this table-top QKD system comprising a simulated satellite, including the source, the synchronization beacon, the self-check and the measurement system, and a receiver unit. Between the simulated satellite and the ground station, a variable attenuation is introduced to simulate link losses.   We regulate the dark count levels, brightness of the source and coincidence window (software) in order match realistic satellite operating conditions. While most of the optical components are representative of the final payload, we use off the shelf  components for control, detection and timing electronics. One critical difference with the payload is the use of a long single mode fiber which introduces polarization rotation that needs to be corrected using additional wave plates. The performances of the detection systems are individually verified using light from a laser (785 nm and 850 nm) coupled from the source side to the receiver systems. The total transmission through the receiver unit from the fibre until the detectors is approximately 83\%. The main limitation is the reflection on the dichroic mirror with a reflection efficiency of 92\%. The other optical efficiencies exceed 98\% for all components.

Initially, we measure the source performance, such as the brightness, QBER heralding efficiency etc, without introducing any losses. With these measurement results as initial conditions the QKD performance model is used to predict the behaviour of these parameter and the final secret key rate for different link losses. For each channel loss introduced by the variable optical attenuator, timestamps are obtained at the simulated satellite and the receiver station. The timestamps are then fed into the QKD software stack in order to get the secret keys along with the other parameters. Figure \ref{fig:TTD} compares the observed parameters with the prediction from the model. Behaviour of heralding efficiency, coincidence, quantum bit error rate and secret key rate with the introduced channel loss are given. It can be observed that even after 40 dB of loss, the QBER is well below the classical limit and it can produce secret keys. 

After the validation of the performance of the system for a given link loss we have simulated a satellite pass in which the loss varies dynamically. Figure \ref{fig:pass} shows the performance of the QKD system over a simulated satellite pass. At time 0, the satellite is at its closest to the ground station and this will contribute to lower QBER and higher sifted key rate. After the error correction and privacy amplification, we obtain 26 Kbits (101 256-bit AES keys) of secret keys over a single pass. With the finite key assumption this corresponds to 13.6 Kbits (53 256-bit AES keys) of secret keys for a single pass.

\subsection{Development strategy}
Spacecraft instrumentation and spacecraft in general are typically developed as a series of models, each used to verify part or all of the requirements posed on the system, culminating in the Flight Model (FM) that is launched into Space. We follow a similar path, starting from a general proof-of-concept table-top demonstration iterated in the previous section. The first approximately to-scale model of the opto-mechanical sub-system of the payload, called the Functional Model (FM), aimed to demonstrate full performance levels in a miniaturized form factor and is intended to be used in end-to-end tests in the lab or using a free-space link. A subsequent model, called the Engineering Model (EM), should also demonstrate functionality after initial environmental tests and confirm the final iteration of the optical layout. A full space qualification campaign is planned for the Qualification Model (QM), which will be used to qualify the design through extensive functional and environmental testing, prior to the Flight Model build and launch. Here, we discuss first results from the functional model build and test and progress on the next models. 

\subsubsection{Functional model}
\begin{figure}[!h]
    \centering
    \includegraphics[scale=0.4]{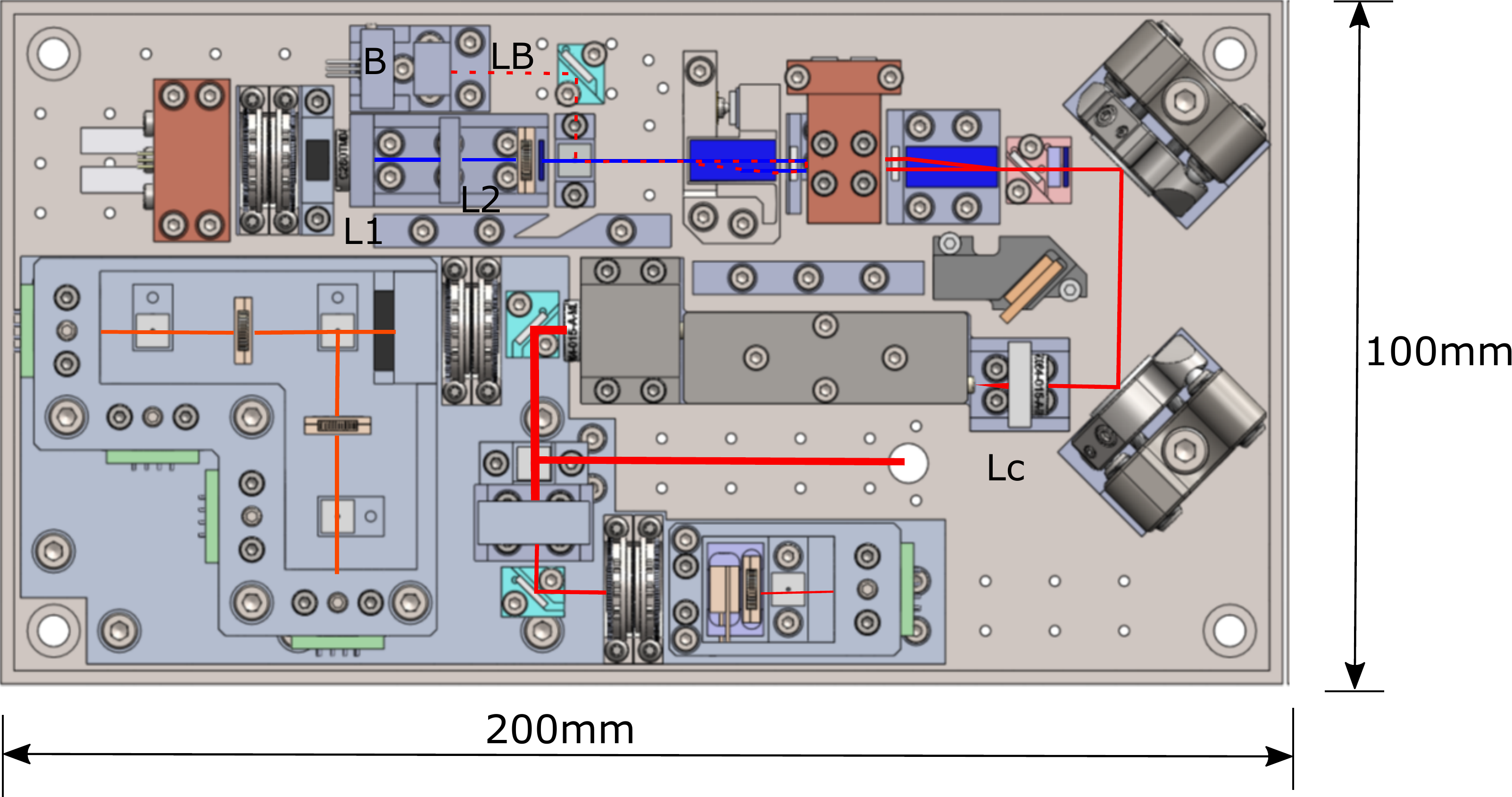}
    \caption{Design of the functional model  }
    \label{fig:fun}
\end{figure}

The design of the functional model is given in Figure \ref{fig:fun}. The main difference between the functional model and the planned final design is that it uses mirrors to steer the SPDC light beam to couple it into a single mode fibre. The SMF is a short piece of fiber kept in an \lq S\rq  shape in order to avoid the propagation of cladding modes. In this model we use a single lens before the fibre. L2 and Lc are moved by external translation stages in the direction of the beam propagation ($\pm z$) to focus the pump for collection. The beacon laser mount (B) is moved in z direction while the beacon lens (LB) is moved in X and Y directions (perpendicular to the beam path) to couple the beacon into the fibre. All the lenses and mounts are secured via epoxy after obtaining the desired performance. 

\begin{figure}[!h]
    \centering
    \includegraphics[scale=.8]{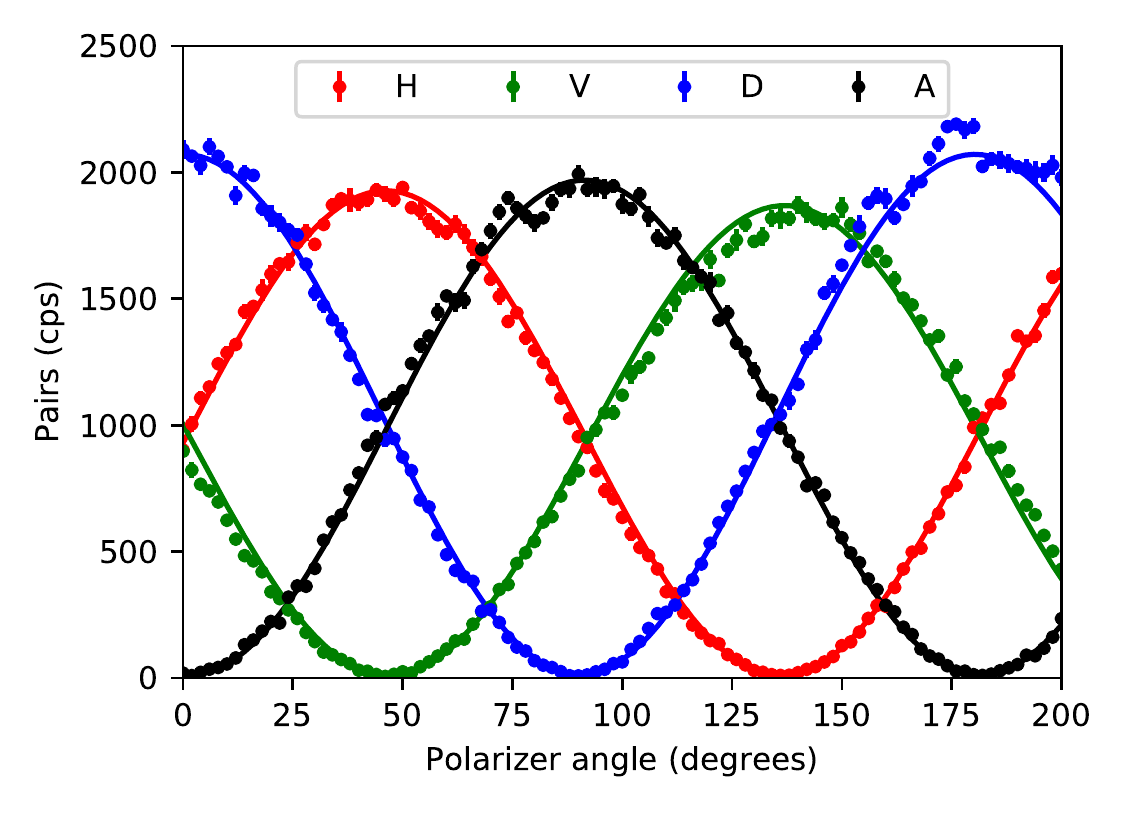}
    \caption{Polarization correlation of the  entangled photons generated in the functional model }
    \label{fig:fumopol}
\end{figure}

After the integration of the source we observed a detected brightness of 600k counts/s/mw with signal and idler heralding efficiencies of 23.5\% and 28\%, respectively. This corresponds to a source brightness of 9M counts/s/mW. With 40 mW of maximum achievable pump power we can easily meet the required brightness of 25-70 MCPs.  The polarization correlation is given in \Cref{fig:fumopol}. The average visibility of the curves is 98.6\% which corresponds to a source QBER of only 0.6\%.  Further integration of the full unit and functionality tests are currently being undertaken.  

\subsubsection{Thermal and vibration tests}

The satellite payload needs to be resilient against random vibrations, thermal gradients and thermal cycles. As alignment tolerances for coupling the light into a single mode fibre are tight, a critical concern is misalignment of the system. We are currently developing a test unit representative of the fiber coupling system of the payload in order to test under different conditions. Once the unit has passed the vibration and thermal tests we will be moving to the final engineering model design. 

\subsubsection{Engineering model}
With the input from the thermal and vibration tests, an engineering model will be developed. This will be a fully functional quantum payload with optical, electronic and software systems. The engineering model will undergo thermal-vacuum testing as well as vibration testing.  

\section{Conclusions}
SpooQy-1 demonstrated the feasibility of operating quantum entanglement sources on board CubeSats and after 600 days of operation, the satellite continuous to produce and detect high quality entangled photons. The effect of the radiation environment on the dark counts of the avalanche photo diodes is measured which is a useful insight for designing next generation quantum payloads. For satellite to ground QKD, requirements are derived from a detailed link model which computes the performance of the system for a given entangled photon source, detectors and timestamping hardware. The model is validated through a table top experiment for QKD. Early development models of the next generation quantum payload have demonstrated performance levels suitable for space-to-ground quantum key distribution and long range entanglement distribution. The next development models will be built and tested this year, with a realistic launch window of 2022-2023. The small volume of 2-4L allows for cost-effective deployment on a wide range of spacecraft, enabling nodes in a quantum network to be deployed in Space.

\acknowledgments 
This research is supported by the National Research Foundation, Singapore under its Central Gap Fund (NRF2018NRF-CG001-001)  

\appendix
\section{Key rate model inputs}
\begin{table}
\caption{\textbf{Inputs for space-to-ground key rate calculations}}
\label{tab:Inputs}
\begin{tabular}{|l|l|c|c|l|}
\hline
\multicolumn{1}{|c|}{\textbf{\begin{tabular}[c]{@{}c@{}}Payload \\ subsystem\end{tabular}}} & \multicolumn{1}{c|}{\textbf{Model inputs}} & \textbf{Unit} & \textbf{Value} & \multicolumn{1}{c|}{\textbf{Reference}} \\ \hline
\begin{tabular}[c]{@{}l@{}}Entangled photon\\  source\end{tabular} & Pair production rate & Mcps & 1-100 & this is a variable input \\ \hline
 & Visibility & \% & 98 & 99\% achieved in the lab \\ \hline
 & Wavelength & nm & 780/842 & \begin{tabular}[c]{@{}l@{}}typical SPDC \\ emission wavelength\end{tabular} \\ \hline
 & Bandwidth & nm & 10nm & 5nm achieved in lab \\ \hline
Detection setups & Efficiency & \% & 0.25 & \begin{tabular}[c]{@{}l@{}}Total detection efficiency \\ or pair-to-singles ratio, \\ 25\% achieved in the lab\end{tabular} \\ \hline
 & Dark counts & cps & 100000/500 & \begin{tabular}[c]{@{}l@{}}SPOOQY-1 results /\\  lab-based numbers (per detector)\end{tabular} \\ \hline
 & Dead time & ns & 50 & \textless{}50ns achieved in lab \\ \hline
 & Jitter & ps & 320 & \begin{tabular}[c]{@{}l@{}}Measured for single photon\\  detectors on SPOOQY-1\end{tabular} \\ \hline
 & \begin{tabular}[c]{@{}l@{}}After-pulsing\\  probability\end{tabular} & \% & \textless{}5 & \begin{tabular}[c]{@{}l@{}}1\% achieved in lab, \\ depends on temperature\end{tabular} \\ \hline
Optical link & Transmitter aperture & m & 0.09 & \begin{tabular}[c]{@{}l@{}}Realistic aperture size \\ for a nanosatellite\end{tabular}  \\ \hline
 & Receiver aperture & m & 0.6 & \begin{tabular}[c]{@{}l@{}}Optimum aperture to \\ cost ratio for \\ PlaneWave telescopes\end{tabular} \\ \hline
 & Beam quality & M2 & 1.6 & \begin{tabular}[c]{@{}l@{}}Fundamental limit is 1.4 \\ due to diffraction\end{tabular} \\ \hline
 & \begin{tabular}[c]{@{}l@{}}Atmospheric\\  attenuation\end{tabular} & dB & 3 & \begin{tabular}[c]{@{}l@{}}3 at zenith from {[}bourgoin{]}, \\ scaled with path length \\ for lower elevation angles\end{tabular} \\ \hline
 & Pointing jitter & microrad & 5 & \begin{tabular}[c]{@{}l@{}}1.2 microrad demonstrated \cite{liao2017satellite}\\ on (larger) Micius satellite\end{tabular} \\ \hline
 & Efficiency & \% & 50 & \begin{tabular}[c]{@{}l@{}}Estimated based on\\ reflectivity and number\\  of optical surfaces\end{tabular} \\ \hline
 & Background counts & cps & 1300 & \begin{tabular}[c]{@{}l@{}}Measured with representative\\  setup in Singapore\end{tabular} \\ \hline
Protocol & \begin{tabular}[c]{@{}l@{}}Error correction\\  efficiency\end{tabular} & - & 1.1 & This depends on the protocol  \\ \hline
 & \begin{tabular}[c]{@{}l@{}}Basis reconciliation\\  factor\end{tabular} & - & 0.5 & \begin{tabular}[c]{@{}l@{}}Un-biased basis splitting\\  in detection setup\end{tabular} \\ \hline
 & \begin{tabular}[c]{@{}l@{}}Coincidence time\\  window\end{tabular} & ns & 1 & \begin{tabular}[c]{@{}l@{}}Lowest achievable\\  with given detector jitter\end{tabular} \\ \hline
Orbit & Altitude & km & 500 & \begin{tabular}[c]{@{}l@{}}Sun-synchronous orbit\\ at 500km means daily passes \cite{liao2017satellite}\end{tabular} \\ \hline
 & \begin{tabular}[c]{@{}l@{}}Maximum elevation\\  angle during pass\end{tabular} & deg & 90 & pass directly over the GS \\ \hline
\end{tabular}
\end{table}
 \pagebreak

\bibliography{report} 
\bibliographystyle{spiebib} 

\end{document}